\documentclass[
 aip,
 jmp,
 amsmath,amssymb,
 reprint,
]{revtex4-2}

\usepackage{graphicx}
\usepackage{dcolumn}
\usepackage{bm}
\usepackage{ulem} 
\usepackage{xcolor}
\usepackage{upgreek}
\usepackage[a4paper, margin=1in]{geometry}
\usepackage{hyperref}

\usepackage{xr-hyper}

\renewcommand{\figurename}{\textbf{Figure}}

\begin{document}

\preprint{AIP/123-QED}

\title{Engineering photomagnetism in collinear van der Waals antiferromagnets}


\author{M.~Na$^{*}$}
\affiliation{Radboud University, Institute for Molecules and Materials, 6525 AJ Nijmegen, The Netherlands}

\author{V.~Radovskaia$^{*}$}
\email{viktoriia.radovskaia@ru.nl}
\affiliation{Radboud University, Institute for Molecules and Materials, 6525 AJ Nijmegen, The Netherlands}

\author{D.~Khusyainov}%
\affiliation{\mbox{Radboud University, Institute for Molecules and Materials, 6525 AJ Nijmegen, The Netherlands}}

\author{P.~Kim}%
\affiliation{\mbox{Radboud University, Institute for Molecules and Materials, 6525 AJ Nijmegen, The Netherlands}}

\author{K.~Mukhuti}%
\affiliation{\mbox{Radboud University, Institute for Molecules and Materials, 6525 AJ Nijmegen, The Netherlands}}
\affiliation{\mbox{HFML-FELIX, Toernooiveld 7, 6525 ED Nijmegen, the Netherlands}}

\author{P. C. M.~Christianen}%
\affiliation{\mbox{Radboud University, Institute for Molecules and Materials, 6525 AJ Nijmegen, The Netherlands}}
\affiliation{\mbox{HFML-FELIX, Toernooiveld 7, 6525 ED Nijmegen, the Netherlands}}

\author{E.~Kochetkova}%
\affiliation{Institute of Physics, University of Amsterdam, Science Park 904, Amsterdam 1098 XH, The Netherlands}

\author{A.~Isaeva}%
\affiliation{Institute of Physics, University of Amsterdam, Science Park 904, Amsterdam 1098 XH, The Netherlands}
\affiliation{\mbox{Department of Physics, Technische Universität Dortmund, 44227 Dortmund, Germany}}
\affiliation{\mbox{Research Center Future Energy Materials and Systems, 44227 Dortmund, Germany}}

\author{A.~de Visser}
\affiliation{Institute of Physics, University of Amsterdam, Science Park 904, Amsterdam 1098 XH, The Netherlands}

\author{D.~Pashov}
\affiliation{\mbox{King’s College London, Theory and Simulation of Condensed Matter, The Strand, WC2R 2LS London, UK}}  

\author{M.~V.~Schilfgaarde}%
\affiliation{Materials, Chemical and Computational Science Directorate, National Laboratory of the Rockies, Golden, CO, 80401 USA}

\author{E. H. T.~Teo}%
\affiliation{\mbox{School of Electrical and Electronic Engineering, Nanyang Technological University, Singapore 639798, Singapore}}
\affiliation{\mbox{School of Materials Science and Engineering, Nanyang Technological University, Singapore 639798, Singapore}}

\author{A.~Chaturvedi}%
\affiliation{\mbox{Temasek Laboratories, Nanyang Technological University, 50 Nanyang Drive, 637553, Singapore}}

\author{S.~Acharya}%
\affiliation{Materials, Chemical and Computational Science Directorate, National Laboratory of the Rockies, Golden, CO, 80401 USA}

\author{Th.~Rasing}%
\affiliation{\mbox{Radboud University, Institute for Molecules and Materials, 6525 AJ Nijmegen, The Netherlands}}

\author{A.~V.~Kimel}%
\affiliation{\mbox{Radboud University, Institute for Molecules and Materials, 6525 AJ Nijmegen, The Netherlands}}

\author{D.~Afanasiev}
 \email{d.afanasiev@science.ru.nl}
\affiliation{\mbox{Radboud University, Institute for Molecules and Materials, 6525 AJ Nijmegen, The Netherlands}}

\begin{abstract}
Achieving efficient ultrafast optical control of antiferromagnetic spin dynamics is a central goal for next-generation high-speed THz spintronic and magnonic devices. Resonant optical pumping of crystal-field–split \textit{d–d} orbital multiplets in magnetic TM ions directly modulates exchange and spin–orbit interactions, inducing large-amplitude coherent spin precession. However, such effects are limited to a handful of systems and there is no general strategy to enhance \textit{d–d} photomagnetism in antiferromagnets. Here, we demonstrate the engineering of photomagnetism via TM-ion doping in collinear van der Waals antiferromagnets. In Mn$_{1-x}$Ni$_x$PS$_3$, small amounts of Ni$^{2+}$ activate a strong photomagnetic response while largely preserving the Néel ground state. Even 10\% Ni boosts the response by more than an order of magnitude compared to pure MnPS$_3$, with resonant pumping of Ni$^{2+}$ \textit{d–d} transitions driving large-amplitude coherent spin precession and providing helicity-dependent phase control. Tuning the pump energy across the full Mn$_{1-x}$Ni$_x$PS$_3$ composition range shows that Ni excitations remain effective across competing Néel and zig-zag antiferromagnetic states while supporting tunable-frequency coherent spin precession. These results establish TM-ion doping as a versatile strategy to harness orbital multiplet excitations for ultrafast, low-dissipation spin control in van der Waals antiferromagnets.

\end{abstract}

\maketitle

\noindent\textsuperscript{*}These authors contributed equally.

\newpage
\section{Introduction}
Photomagnetism, the direct control of magnetic states through resonant light-matter interactions~\cite{enz1969} offers a route toward ultrafast and energy-efficient spin manipulation beyond the limits of magnetic-field-based approaches \cite{kampfrath2013resonant,tesavrova2013experimental,satoh2012directional}. To date, photomagnetism has been observed across a wide range of material platforms, from single-molecule magnets~\cite{bleuzen2009photomagnetism, zakrzewski2024optical}, to ferro- and antiferromagnetically (AFM) ordered solids~\cite{stupakiewicz2017, Baierl2016, disa2023photo}. Unlike optomagnetic effects, such as the inverse Faraday \cite{kimel2005ultrafast} or Cotton-Mouton effects \cite{kalashnikova2007impulsive}, which are non-dissipative and, in principle, do not require photon absorption, photomagnetic effects rely explicitly on absorption and are therefore expected to be highly efficient in controlling spins~\cite{kirilyukUltrafast2010}. Photon absorption promotes specific electronic or phononic excitations that modify orbital occupations and, consequently, magnetic anisotropy and exchange interactions \cite{afanasiev2021ultrafast,afanasiev2021,mikhaylovskiy2020,ron2020ultrafast,nova2017effective,disa2023photo}. Such resonant control provides a fundamentally different paradigm from laser-induced heating, opening tantalizing prospects for coherent manipulation of spins.


This capability is particularly critical in collinear antiferromagnets. Their vanishing net magnetization precludes conventional magnetic-field control, while their intrinsic terahertz spin dynamics demand optical-bandwidth excitation with spin-precession amplitudes sufficient for switching \cite{Nemec2018,satoh2012directional,satoh2015writing,kampfrath2011coherent}. 
Among available strategies, the resonant optical pumping of crystal-field split orbital multiplets, such as localized \textit{d-d} and \textit{f-f} transitions in transition-metal (3$d$) and rare-earth (4$f$) antiferromagnets, has emerged as a promising route \cite{stupakiewicz2017,Schlauderer2019,Baierl2016,stupakiewicz2019,afanasiev2021}. 
In magnetic transition-metal (TM) ions, these transitions are nominally parity- (Laporte) and spin-forbidden in centrosymmetric environments and acquire oscillator strength only through symmetry breaking mechanisms~\cite{marciniak2021vibrational,lohr1972spin,marciniak2021vibrational}, for example via vibronic coupling, local lattice distortions, or hybridization with ligands. Nevertheless, recent progress has shown that resonant femtosecond pumping of these states can generate exceptionally strong photomagnetic responses, despite low absorption. In particular, resonant \textit{d-d} photoexcitation can drive large-amplitude spin precession in AFMs by transiently modifying exchange interactions~\cite{mikhaylovskiy2020}, anisotropy~\cite{Baierl2016, afanasiev2021, fitzky2021, matthiesen2023}, and has even enabled switching between metastable states in ferrimagnets~\cite{stupakiewicz2017, stupakiewicz2019}. 
Yet, a predictive framework for engineering a coherent photomagnetic response in collinear antiferromagnets remains lacking. Understanding why certain orbital transitions drive giant coherent spin motion whereas others -- even within the same magnetic ion -- remain ineffective is therefore a central and unresolved challenge~\cite{iida2011spectral, mikhaylovskiy2020, afanasiev2021,belvin2021,mertens2023ultrafast}. 

In this work, we aim to elucidate the fundamental principles governing the efficacy of \textit{d-d} orbital multiplet excitations by engineering the coherent response in alloys of NiPS$_3$ and MnPS$_3$. These isostructural compounds belong to the TM-thiophosphate family of van der Waals antiferromagnets~\cite{chittariElectronic2016, basnetUnderstanding2024}, renowned for their exceptionally strong light-matter interactions, pronounced excitonic correlations, and, most importantly, their ability to launch large-amplitude coherent spin dynamics through resonant pumping of \textit{d-d} orbital transitions~\cite{Baierl2016, afanasiev2021, matthiesen2023,shan2021giant,lee2024,Toyoda2024,belvin2021}. These two TM ions host different orbital manifolds, spin multiplicities, and spin-orbit couplings, which lead to different antiferromagnetic ground states \cite{dirnberger2022spin,lebedev2025ultranarrow,kang2020,basnetUnderstanding2024}. By partially substituting Mn with Ni in the Mn$_{1-x}$Ni$_x$PS$_3$ series, we get an opportunity to tune the magnetic ground state between the Néel order of MnPS$_3$ and the zig-zag order of NiPS$_3$. This substitution also enables a direct comparison of the photomagnetic responses of Mn$^{2+}$ and Ni$^{2+}$ ions, achieved by scanning the pump photon energy across their respective \textit{d–d} multiplet resonances and measuring the amplitude of the resulting collective spin dynamics. Strikingly, at 10\% Ni -- where the Néel ground state of MnPS$_3$ is preserved -- the spin precession driven by Ni$^{2+}$ multiplets is nearly 15× larger than that generated by Mn$^{2+}$ multiplets, while also providing helicity-dependent phase control, absent in pure MnPS$_3$. Moreover, it is the $^3$A$_{1g}$ multiplet excitation -- a minor feature in the absorption spectrum of NiPS$_3$ -- that emerges as the most efficient driver of spin dynamics, surpassing all Mn$^{2+}$ multiplets and the more optically absorptive Ni$^{2+}$ multiplets.

Supported by quasiparticle self-consistent $G\hat{W}$ (QS$G\hat{W}$) theory~\cite{Cunningham2023, questaal_paper,acharyaTheoryColorsStrongly2023} -- which allows direct visualization of the excited-state orbital wavefunctions -- we identify the key factors that govern the photomagnetic response of different multiplet excitations: (i) although Mn$^{2+}$ transitions involve spin-flip and directly couple to spins, their weak oscillator strength makes them far less effective than the optically stronger, spin-conserving Ni$^{2+}$ resonances; (ii) the strongest drivers of spin precession are those excitations with highly localized wavefunctions that lie away from the fundamental bandgap and minimize hybridization with the surrounding sulfur ligand field; and (iii) the lowest-energy Ni$^{2+}$ $^3A_{1g}$ state carries substantial orbital angular momentum, and couples more efficiently to the spin degree of freedom than other Ni$^{2+}$ multiplet excitations. Optical pumping of this resonance, therefore, generates coherent spin-precessions with amplitudes more than an order of magnitude larger than those observed via Mn$^{2+}$ excitations. Altogether, our findings establish the orbital degree of freedom as a powerful driver of spin dynamics and demonstrate that strategic ion substitution offers a versatile route for shaping the photomagnetic response and controlling collective excitations in collinear antiferromagnets.

\section{Tuning magnetism in the mixed-compounds}
\begin{figure*}[!t]
    \centering
    \includegraphics[width = 1\textwidth]{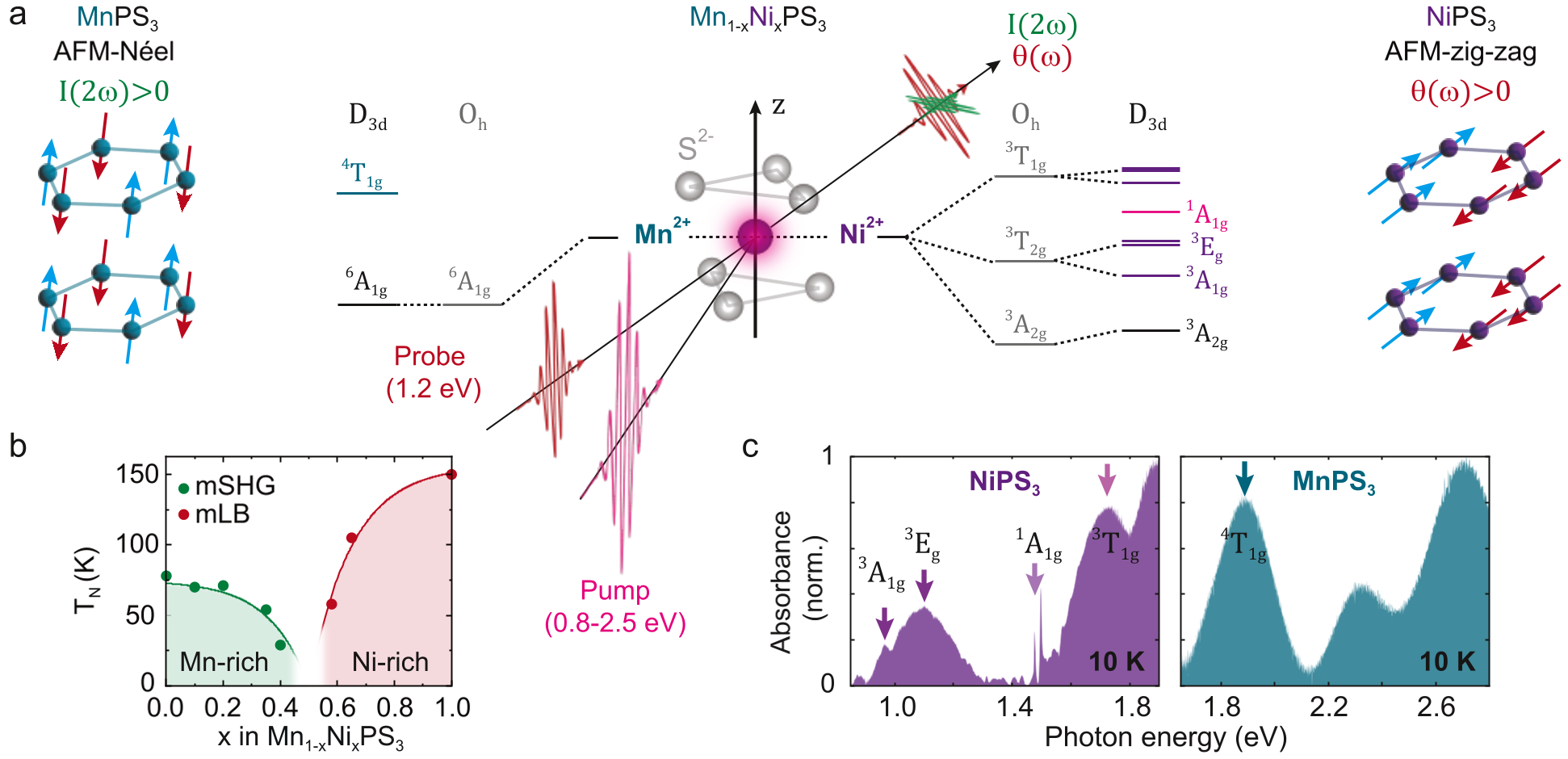}
    \caption{
    \textbf{Tuning the antiferromagnetic ground state and photomagnetic properties of Mn$_{1-x}$Ni$_x$PS$_3$.} \textbf{(a)} The transition-metal (TM) ions form a hexagonal lattice and are surrounded by trigonally-distorted S$_6$ octahedral cages. The Tanabe-Sugano diagram of Mn$^{2+}$ and Ni$^{2+}$ ions in the octahedral ($O_h$) crystal field with a trigonal ($D_{3d}$) distortion is shown in the center, along with spin-flip excited states. MnPS$_3$ (left) favours a N\'eel (N)-type antiferromagnetic (AFM) order with the N\'eel vector out-of-plane. In contrast, the larger spin-orbit coupling leads to an in-plane anisotropy for NiPS$_3$ and zig-zag (ZZ) AFM order (right). In the pump-probe experiment, the laser fundamental (1.2~eV) is detuned from the multiplet resonances and is used as a probe. The pump pulse, with an incident fluence of 6~mJ/cm$^2$, is tuned from 0.8 to 2.4~eV across multiple multiplet resonances. \textbf{(b)} The ground state magnetic phase diagram of Mn$_{1-x}$Ni$_x$PS$_3$ mapped by symmetry-selective magneto-optical probes. The spatial inversion symmetry breaking in the N-type AFM order gives magnetic second-harmonic generation (mSHG, green) and the ZZ-type order gives magnetic linear birefringence (mLB, red); these probes are used to determine the onset of the AFM order at the N\'eel temperature ($T_\text{N}$). The Mn (Ni)-rich compounds have an N (ZZ)-type AFM ground state reflective of the pure compound, while long-range magnetic order is not observed for $x = 0.5$. \textbf{(c)} Optical absorbance of pure NiPS$_3$ and MnPS$_3$ compounds measured at 10~K, showing the below-bandgap multiplet resonances.
    }
    \label{Fig: fig1}
\end{figure*}

The Mn$_{1-x}$Ni$_{x}$PS$_3$ samples were grown via chemical vapor transport~\cite{Susner2017} and characterized with energy-dispersive X-ray spectroscopy (EDS) to confirm a homogeneous elemental distribution (see Supplementary Sec. 1). In their pure forms, MnPS$_3$ and NiPS$_3$ both adopt a monoclinic crystal structure that can be described as a hexagonal lattice of TM-S$_6$ distorted octahedra with a phosphorus dimer in the center. Despite their structural similarities, the magnetic properties of MnPS$_3$ and NiPS$_3$ differ due to the distinct orbital occupation of the TM ions and the resulting spin configuration ~\cite{joyMagnetism1992, basnetHighly2021, basnetUnderstanding2024}. MnPS$_3$ adopts a N\'eel (N)-type order while NiPS$_3$ exhibits a zigzag (ZZ)-type order, as shown in the left and right schematics of Fig.~\ref{Fig: fig1}a. The N-type AFM ground state breaks spatial inversion symmetry, resulting in a significant second-order susceptibility and the emergence of magnetic second-harmonic generation (mSHG) below $T_\text{N} = 78$~K~\cite{chu2020}. The ZZ-type AFM ground state in NiPS$_3$ preserves spatial inversion symmetry; however, the formation of extended ferromagnetic chains induces strong optical anisotropy, which manifests as magnetic linear birefringence (mLB) below $T_\text{N}=155$~K~\cite{zhangObservation2021, khusyainov2023, rybak2024}. Furthermore, the N\'eel vector is out-of-plane in MnPS$_3$, which is known for the weakest single-ion anisotropy in the family ~\cite{okuda1986}. In contrast, in NiPS$_3$ a significantly larger single-ion anisotropy aligns the N\'eel vector in-plane~\cite{Wildes2006, lanccon2018}. 

To characterize the AFM ground state of the mixed Mn$_{1-x}$Ni$_x$PS$_3$ compounds we employ the two symmetry-selective magneto-optical probes mSHG and mLB. In Fig.~\ref{Fig: fig1}b, we combine the temperature-dependence of mSHG (green) and mLB (red) to generate an AFM ground state phase diagram for Mn$_{1-x}$Ni$_x$PS$_3$, as a function of temperature and the Ni fraction $x$. The experimental methods and examples of the mSHG and mLB temperature dependencies are shown in Supplementary Sec. 2. We observe that only Mn-rich samples ($x < 0.5$) exhibit mSHG, while mLB only occurs in Ni-rich samples ($x > 0.5$). This trend suggests that the quantitatively dominant TM ion dictates the type of AFM order, with Mn-rich compositions favouring the N-type and Ni-rich compositions favour the ZZ-type ground state. Further, we observe a suppression of $T_\text{N}$ upon mixing. The suppression reflects the competition between AFM nearest-neighbour exchange preferred by Mn ions and the FM nearest-neighbour exchange preferred by Ni ions, which destabilizes long-range magnetic order and leads to the disappearance of both the mLB and mSHG signals at half-mixing ($x = 0.5$). This observation is in line with previous studies indicating that compositions near $x=0.5$ may host a spin-glass ground state~\cite{luEvolution2022}. 

\section{Optically-induced collective spin dynamics}
Having established the AFM ground state, we turn to the photomagnetically-induced coherent spin dynamics. In the ground state of both MnPS$_3$ ($^6$A$_{1g}$) and NiPS$_3$ ($^3A_{2g}$), the orbital angular momentum is largely quenched by the octahedral crystal field. Even so, resonant excitation to higher-energy orbital multiplets -- $^3T_{2g}$ ($\approx$1~eV), $^1A_{1g}$ ($\approx$1.5~eV) in Ni$^{2+}$ and $^4T_{1g}$ ($\approx$1.9~eV) in Mn$^{2+}$ -- has been shown to drive collective spin precession in their respective pure compounds~\cite{afanasiev2021, belvin2021, matthiesen2023, allington2025distinct}. These resonances, shown schematically in Fig.~\ref{Fig: fig1}a appear as distinct absorption peaks in the below-bandgap optical absorption spectra (Fig.~\ref{Fig: fig1}c) and can be accessed by tuning the pump photon energy $h\nu$. Importantly, the Mn and Ni multiplet excitations are well separated in energy and do not spectrally overlap, allowing for a direct and unambiguous comparison of their efficiency in driving coherent magnons. 

\begin{figure*}[!t]
    \centering
    \includegraphics{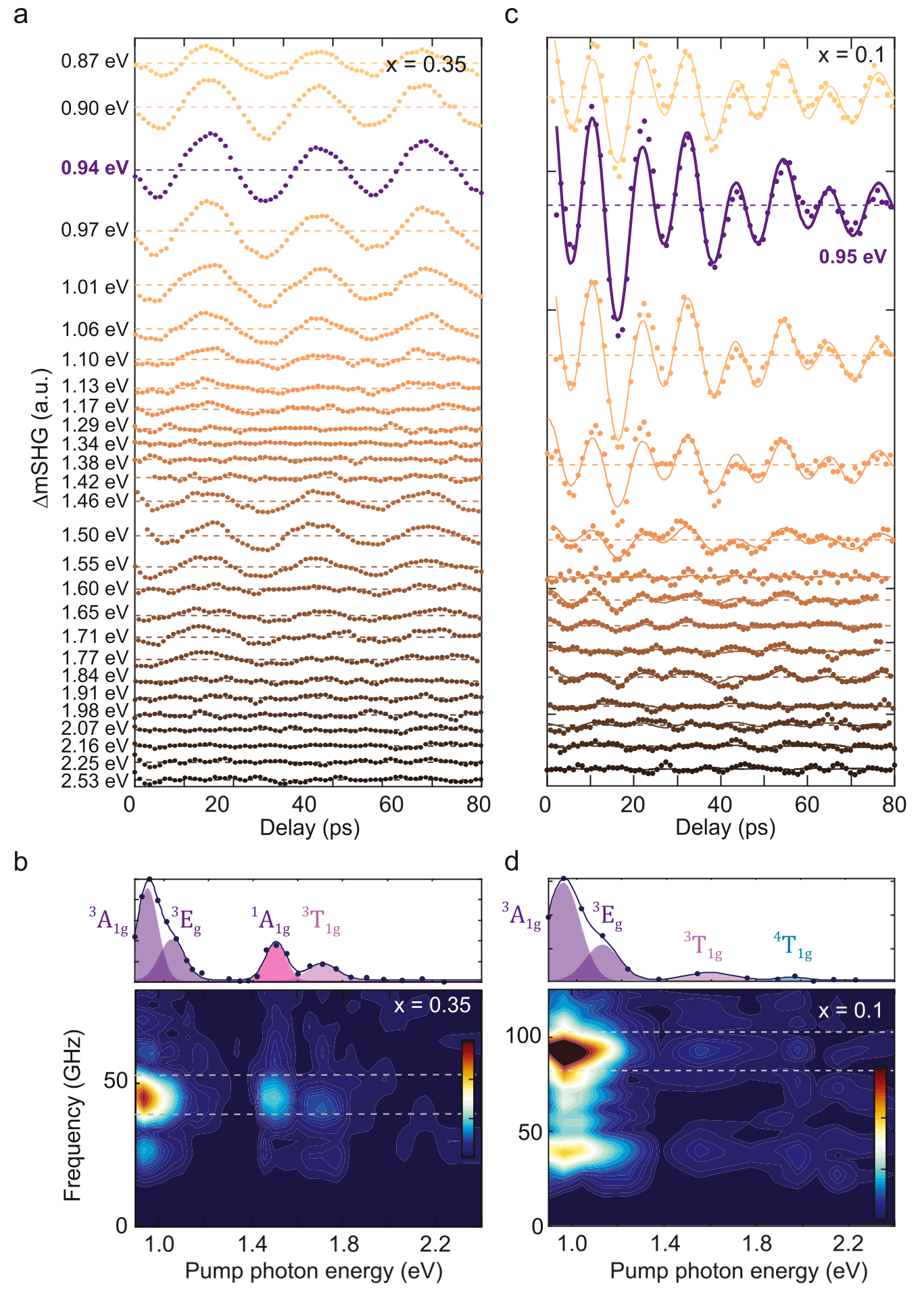} 
    \caption{
    \textbf{Launching coherent spin dynamics via resonant multiplet excitation in Mn$_{1-x}$Ni$_{x}$PS$_3$.} \textbf{(a, c)} The pump-induced change of the magnetic second harmonic signal ($\Delta$mSHG) as a function of pump-probe delay measured for $x = 0.35$ and $x = 0.1$ at $T = 10$~K, respectively. The trace at 0.94 (0.95)~eV is highlighted. \textbf{(b, d)} The frequency spectrum of the coherent dynamics as a function of pump photon energy in the $x = 0.35$ and $x = 0.1$ samples. Colors indicate $\Delta$mSHG intensity in linear scale. In panel b, clear resonances are observed at $E_\text{Ni} = [0.925, 0.97, 1.498, 1.708]$~eV, corresponding to multiplet resonances in Ni$^{2+}$ in the trigonally distorted octahedra. In panel d, a clear resonance is observed at 0.95~eV, and smaller resonances at 1.55, and 1.98~eV, corresponding to multiplet excitations in Ni$^{2+}$ (purple) and Mn$^{2+}$ (turquoise). The intensity between the gray dashed lines is integrated and shown for the $x = 0.35$ (left) and $x = 0.1$ (right) samples and fit with four Gaussian peaks.
    }
    \label{Fig: fig2}
\end{figure*}

We begin with the Mn-rich sample Mn$_{0.65}$Ni$_{0.35}$PS$_3$ ($x = 0.35$) where the Mn to Ni ratio is approximately 2:1. This ensures substantial contribution from both magnetic ions while preserving a robust N-type ground state. The incident pump fluence is fixed at 6~mJ/cm$^2$, and the pump photon energy is tuned from 0.8 to 2.5~eV to span the full range of Mn and Ni multiplet resonances. Spin dynamics are monitored using transient changes in the magnetic second-harmonic generation ($\Delta$mSHG) in a transmission geometry, with a linearly polarized probe at 1.2 eV -- outside the multiplet resonance range of either ion. The pump-induced dynamics as a function of pump-probe delay is shown in Fig.~\ref{Fig: fig2}a. Here we present only the coherent oscillatory component of the data; for the raw data and the analysis procedure, see Supplementary Sec. 3. The data show that the oscillation amplitude depends strongly on the pump photon energy, with the largest response observed at 0.94~eV (purple).

The Fourier transform of the oscillations (Fig.~\ref{Fig: fig2}b) reveals two modes with frequencies of 45 and 25~GHz. Their magnetic origin is evident from their temperature dependence (see Supplementary Sec. 4). As the temperature increases, a pronounced softening of the largest-amplitude 45~GHz mode towards $T_\text{N} = 52$~K is observed -- characteristic of an antiferromagnetic magnon. To highlight the resonances that most effectively drive spin dynamics, we plot the Fourier spectra of the oscillations as a function of pump photon energy $h\nu$. Here, we extract the intensity of the 45~GHz mode by integrating the intensity between the two white lines in Fig.~\ref{Fig: fig2}b. The spectral weight as a function of $h\nu$ shown at the top of Fig.~\ref{Fig: fig2}b is best fit with four Gaussian curves, which peak at $E_\text{pk} = [0.925(10), 0.98(4), 1.498(10), 1.71(5)]$~eV, closely matching the $^3A_{1g}$, $^3E_{g}$, $^1A_{1g}$ and $^3T_{1g}$ optical absorption peaks of Ni$^{2+}$ ions (see Fig.~\ref{Fig: fig1}c). The two lowest-energy peaks -- $^3A_{1g}$, $^3E_{g}$ -- derive from the doubly-degenerate $^3T_{2g}$ orbital multiplet in an ideal $O_h$ crystal field and are split here by the trigonal distortion of the $S_6$ octahedra.
Remarkably, despite Mn being the dominant magnetic species in the mixture, no indication of photoinduced spin precession is observed when the pump photon energy $h\nu$ is tuned into resonance with the Mn$^{2+}$ $^6A_{1g}$ to $^4T_{1g}$ transition at 1.9 eV. This absence is particularly striking because the Mn$^{2+}$ orbital excitation intrinsically involves a spin-flip process ($\Delta S = 1$), and its' optical excitation is therefore expected to couple directly to the spin sector, potentially launching a large-amplitude coherent spin-wave response. The lack of such a response indicates that in the considered mixture, the Ni$^{2+}$ crystal-field excitations provide the dominant pathway for driving coherent antiferromagnetic spin dynamics.

To highlight the exceptional spin response to the Ni$^{2+}$ multiplet excitations, we performed the same measurements on Mn$_{0.9}$Ni$_{0.1}$PS$_3$ ($x = 0.1$), see Fig.~\ref{Fig: fig2}c. This composition is strongly Mn-rich with a 9:1 Mn:Ni ratio, nearly five times higher than in Mn$_{0.65}$Ni$_{0.35}$PS$_3$, and has a robust N-type antiferromagnetic ground state (see Fig.\ref{Fig: fig1}b). Similarly to Mn$_{0.65}$Ni$_{0.35}$PS$_3$, the pump-photon-energy dependent Fourier spectrum in Fig.~\ref{Fig: fig2} also reveals two modes with frequency 92~GHz and 39~GHz. The higher-frequency mode is softened relative to the doubly degenerate zone-center magnon in pure MnPS$_3$ at 120 GHz~\cite{Wildes1998}, indicating a reduction of the effective exchange interactions induced by Ni substitution. The appearance of the lower-frequency mode points to a lifting of the magnon degeneracy, which we attribute to the progressive development of in-plane anisotropy promoted by Ni$^{2+}$ ions ~\cite{basnetHighly2021}. As described above, increasing the Ni fraction to $x = 0.35$ results in a further renormalization of the two modes to 45 and 25~GHz, consistent with the progressive modification of exchange interactions and anisotropy.

Figure \ref{Fig: fig2}d reveals that although the enhanced Mn fraction makes the contribution of the spin-flip Mn$^{2+}$ $^4T_{1g}$ multiplet resonance to the excitation of spin precession discernible, its overall impact on the magnetic system remains weak. 
In contrast, photoexcitation of the minority Ni$^{2+}$ multiplets results in the unambiguous excitation of coherent spin dynamics. Even more unexpectedly, despite having the smallest absorption cross section, the $^3A_{1g}$ resonance drives the largest amplitude magnons, even in the $x = 0.1$. In contrast, the $^1A_{1g}$ and $^3T_{1g}$ resonances are only tentatively identified for the $x = 0.1$ sample, reflecting the reduced Ni concentration. Altogether, these observations show that the spin oscillations are photomagnetically driven by the selective excitation of specific multiplet resonances. The absence of any scaling with optical absorption rules out a heat-driven mechanism and identifies the dynamics as fundamentally nonthermal.

The large photomagnetic response of Mn$_{1-x}$Ni$_{x}$PS$_3$ to the weakly-absorbing $^3A_{1g}$ and $^3E_{g}$ resonances could in principle be attributed to the increased pump penetration depth, as the transmitted second-harmonic signal is most sensitive to the back surface of the sample. To ensure the observed photomagnetism is not due to a difference in pump fluence at the back surface, we complemented the $\Delta$mSHG measurements in transmission for the $x = 0.1$ sample with $\Delta$mSHG in reflection and $\Delta$mLB in transmission -- probes that are significantly less sensitive to pump-probe penetration depth mismatch. These complementary probes -- shown in the Supplementary Sec. 5 -- consistently reveal that the Ni$^{2+}$ $^3A_{1g}$ resonance drives magnons with higher amplitude than other multiplet excitations.

To account for the pronounced sensitivity of the collective spin precession to specific Mn$^{2+}$ and Ni$^{2+}$ orbital transitions, it is essential to consider both how photoexcited orbital states couple to the spin system and how efficiently light drives the corresponding \textit{d-d} transitions. We begin by comparing the orbital excitations of Mn$^{2+}$ and Ni$^{2+}$ ions, which exhibit radically different photomagnetic responses. All Mn$^{2+}$ excitations from the high-spin $^6A_{1g}$ ground state are spin-flip in character ($\Delta S = 1$), whereas most Ni$^{2+}$ transitions are spin conserving ($\Delta S = 0$). Although spin-flip processes might be expected to couple more directly to magnetic order, the experiments show the opposite: even at a Mn:Ni ratio of 9:1, the Mn $^4T_{1g}$ resonance drives magnons with amplitudes nearly 15 times smaller than those generated by the weak Ni $^3A_{1g}$ transition. 
The stark contrast therefore cannot be explained by the spin character of the transitions alone, but instead reflects how efficiently the optical field couples to the respective orbital excitations. The weak photomagnetic response of Mn is rather consistent with the suppression of oscillator strength in \textit{d–d} transitions that involve a spin flip and are known to reduce optical coupling by up to two orders of magnitude~\cite{acharyaTheoryColorsStrongly2023,jana2025deconstruction}. Consequently, despite providing a formally direct magnetic pathway, pumping of the Mn excitations is far less effective than that of the optically stronger, spin-conserving Ni resonances, which dominate the observed coherent photomagnetic response.

\begin{figure*}[!t]
    \centering
    \includegraphics[width = 1\textwidth]{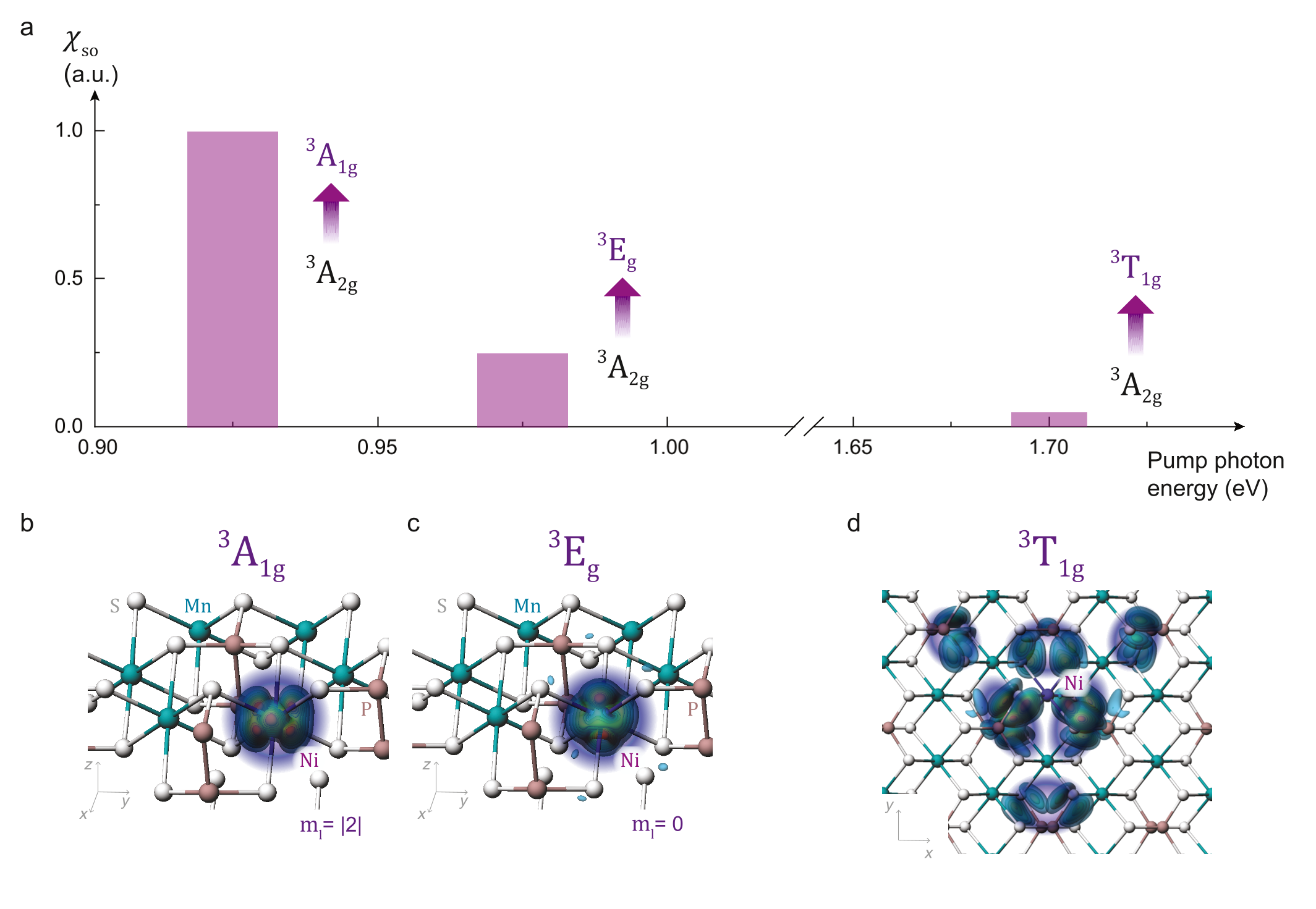} 
    \caption{\textbf{Spin-orbit efficiency of Ni$^{2+}$ excitations in Mn$_{0.75}$Ni$_{0.25}$PS$_3$.}
    \textbf{(a)} The orbital-to-spin coupling strength $\chi_{so}$ of the $^3A_{1g}$, $^3E_{g}$ and $^3T_{1g}$ multiplets, which is calculated as the intensity of the spin dynamics (the amplitude of the corresponding Gaussian peaks in Fig. \ref{Fig: fig2}b) divided by the absorption peaks (the values of absorption at corresponding photon energies in NiPS$_3$ presented in Fig. \ref{Fig: fig1}c) and normalized to 1.
    \textbf{(b-d)} The wavefunction of the excitonic-hole for the (b) $^3A_{1g}$, (c) $^3E_{g}$ and (d) $^3T_{1g}$ multiplets. The Ni and Mn ions are shown in purple and turquoise, respectively. S and P are shown in gray and brown. The $^3A_{1g}$ and $^3E_g$ excitations are localized to the Ni-ion, while the $^3T_{1g}$ state is delocalized over multiple hexagonal plaquette. $m_l = |2|, (0)$ labels the Ni $d$ states onto which the $^3A_{1g}$ ($^3E_g$) excitation is projected.
    }
    \label{Fig: fig3}
\end{figure*}

However, oscillator strength alone cannot account for the large photomagnetic response of the $^3A_{1g}$ excitation, which is a relatively weak absorption feature among Ni multiplet resonances. To highlight the relative efficiency of the photomagnetic orbital-to-spin coupling, we took the maximum amplitude of the driven spin precession in the Mn$_{0.7}$Ni$_{0.3}$PS$_3$ sample (from Fig. \ref{Fig: fig2}b) and normalized it by the corresponding maximum of the absorption spectrum, obtaining a value of orbital-to-spin coupling strength $\chi_{so}$. Plotting these values as a histogram (see\, Fig.\,\ref{Fig: fig3}a) reveals a clear trend, showing that the $^3A_{1g}$ orbital state couples significantly more strongly to the spins compared with the $^3E_{g}$ and $^3T_{1g}$ states, despite its weaker optical absorption.

To understand the origin of these differences, we analyze the excited states using the \textit{ab-initio} many-body perturbation theory (MBPT) framework, QS$G\hat{W}$~\cite{Cunningham2023}, which provides a detailed description of the electronic structure of Mn$_{1-x}$Ni$_x$PS$_3$ (details in Supplementary Sec. 6). Using this approach, we calculated the single-particle electronic structure for $x=0.25$ compositions, approximating the experimental samples, modeled in a 40-atom supercell; see Extended Data Fig.~\ref{Fig: ext}. While the top of the valence bands shows strong hybridization between Mn-\textit{d} and S-\textit{p} states, the conduction bands are predominantly of Ni-\textit{d} character. The occupied Ni-\textit{d} states are located near $-0.75$ eV, as detailed in reference~\cite{jana2025deconstruction}, which provides a comprehensive analysis of the one- and two-particle properties of Ni and Mn compounds individually. Importantly, the multiplet excitations observed in the material are excitonic in nature rather than single-particle transitions.

Using these results, we first analyze the real-space excitonic wavefunctions of the closely spaced $^3A_{1g}$ and $^3E_{g}$ states. Although both originate from the $^3T_{2g}$ level, split by the trigonal distortion of the $S_6$ octahedra in NiPS$_3$~\cite{bandaOptical1986, kim2024}, they exhibit very different photomagnetic orbital-to-spin coupling strength $\chi_{so}$. Figure \ref{Fig: fig3}b-c show that the hole wavefunctions of both states are localized on the Ni ion, with $^3A_{1g}$ lying primarly in the $xy$-plane and $^3E_{g}$ extending along the $z$-axis. The projection of the single-particle valence wavefunctions onto the excitonic hole component, shown in Extended Data Fig. \ref{Fig: ext}, demonstrates that both states have $e_g$ character, corresponding to orbital angular momenta $m_l = |2|$ and 0, largely associated with the $d_{x^2-y^2}$ and $d_{z^2}$ orbitals, respectively. The $^3A_{1g}$ multiplet is more heavily weighted toward $m_l = |2|$, resulting in stronger spin-orbit coupling, whereas $^3E_{g}$ has smaller orbital angular momentum and weaker coupling. Consequently, the $^3A_{1g}$ excitation couples more effectively to the spin sector, consistent with the larger magnon amplitude it drives.

We finally analyze the $^3T_{1g}$ excitation, which, despite having the strongest absorption, exhibits minimal orbital-to-spin coupling strength $\chi_{so}$. Unlike the highly localized $^3A_{1g}$ and $^3E_{g}$ states, $^3T_{1g}$ lies near the Ni band edge at 1.7 eV, indicating weaker excitonic binding and a delocalized wavefunction \cite{jana2025deconstruction,paulina}. As shown in Fig.~\ref{Fig: fig3}d, the $^3T_{1g}$ hole-wavefunction extends across several hexagon plaquettes. Momentum-space decomposition (Extended Data Fig.~\ref{Fig: ext}h) reveals that the excitonic wavefunction is dominated by S-$p$ valence states, which do not produce a change in the Ni orbital angular momentum and therefore couples only weakly to the spin sector. Consequently, the $^3T_{1g}$ resonance is inefficient in driving coherent spin dynamics, consistent with previous reports on orbital pumping in NiPS$_3$, where this excitation produced no appreciable magnon response~\cite{afanasiev2021}.

\section{Phase control of magnons in $\mathbf{Mn}_{1-x}\mathbf{Ni}_{x}\mathbf{PS}_3$}
Having identified the Ni $^3A_{1g}$ excitation as the most efficient trigger of spin dynamics, even at low concentrations, we now examine how it can also be used to control these dynamics. In NiPS$_3$, the amplitude and phase of spin precession driven by the $^3A_{1g}$ multiplet can be controlled by both the helicity of a circularly-polarized pump and the orientation of a linearly-polarized pump \cite{Toyoda2024}. In contrast, MnPS$_3$ shows no reported polarization-dependent control. To test whether the $^3A_{1g}$ multiplet excitation can impart similar control of spin waves in Mn$_{1-x}$Ni$_x$PS$_3$, we study the linear polarization and helicity dependence of the magnon phase as a function of the Ni fraction ($0.1\leq x \leq 0.9$). Spin dynamics are tracked through the pump-induced change in magneto-linear birefringence ($\Delta$mLB), which provides a composition-independent readout even in samples where the equilibrium mLB is absent. 

\begin{figure*}[!t]
    \centering
    \includegraphics[width = 1\textwidth]{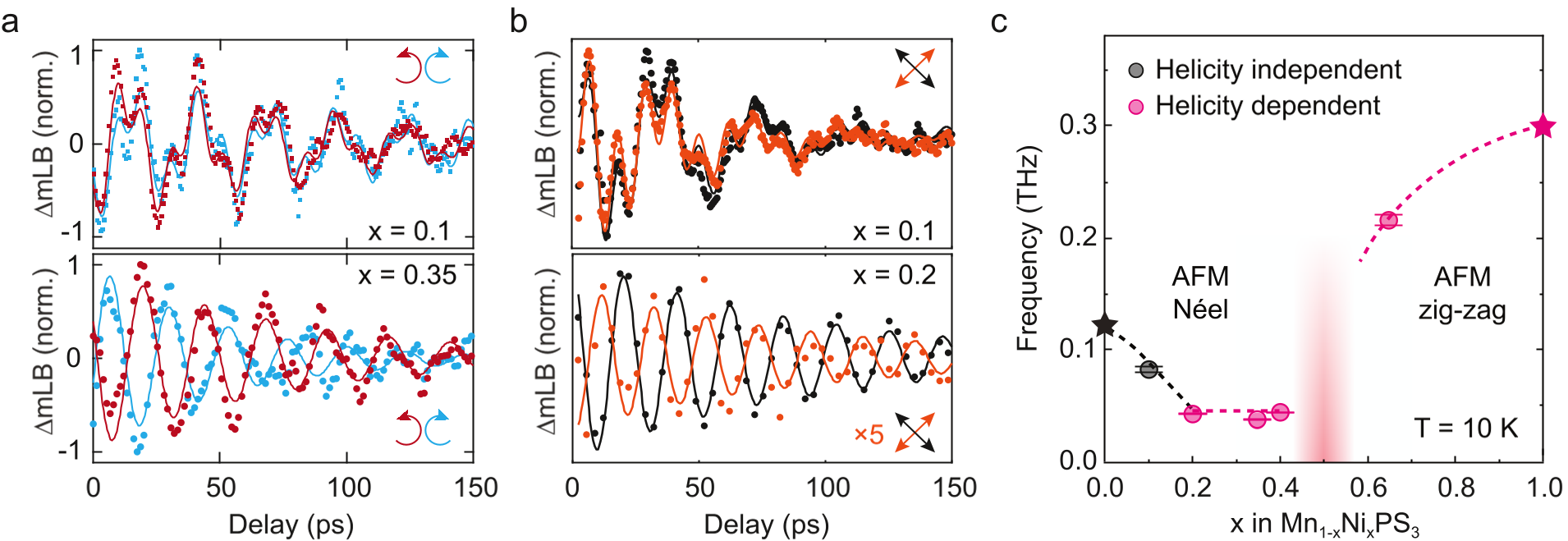} 
    \caption{
    \textbf{Phase control of spin dynamics in Mn$_{1-x}$Ni$_x$PS$_3$.} \textbf{(a)} Helicity dependence of the magnon excitation for the $x = 0.1$ (top) and $x = 0.35$ (bottom) samples at 10~K. \textbf{(b)} Linear polarization dependence of the magnon excitation for the $x = 0.1$, $x = 0.2$ samples, respectively, at 10~K. The orange curve in the bottom panel is vertically scaled by a factor of 5 for visualization purposes. \textbf{(c)} Frequency of the magnons driven by the Ni $^3A_{1g}$ excitation (0.95~eV) as a function of Ni fraction, measured with magnetic linear birefringence. Black (magenta) colors indicate the absence (presence) of phase-flip. The star markers for $x = 0$ and $x=1$ are values reported in literature~\cite{afanasiev2021, matthiesen2023}. 
    }
    \label{Fig: fig4}
\end{figure*}

Fig.~\ref{Fig: fig4}a shows the spin dynamics induced by circularly-polarized pump pulses for the $x=0.1$ and $x=0.35$ samples. We observe no helicity-dependent response in the $x=0.1$ sample; however, with increasing Ni content, a $\pi$ phase shift emerges between opposite helicities -- consistent with reports on pure NiPS$_3$~\cite{afanasiev2021, Toyoda2024}. A similar trend is observed in the linear polarization dependence (Fig.~\ref{Fig: fig4}b): The $x = 0.1$ sample shows no phase and amplitude variation, whereas both appear at $x = 0.2$. More data are presented in Supplementary Sec. 7. This onset coincides with a change in the magnetic anisotropy seen in magnetometry (see Supplementary Sec. 8), in which the N\'eel vector orientation becomes fully in-plane. Figure~\ref{Fig: fig4}c summarizes the magnon frequencies and helicity dependence across the Mn$_{1-x}$Ni$_x$PS$_3$ series, with black (pink) circles corresponding to the absence (presence) of helicity dependence. Except for $x = 0.1$, the $^3A_{1g}$ resonance enables both efficient magnon generation and optical control over magnon amplitude and phase -- even in Mn-rich compositions where the antiferromagnetic ground state is N-type. 

\section{Conclusion}
In this work, we have outlined a new strategy for engineering multiplet-driven photomagnetism in collinear AFMs. We demonstrate that substituting TM ions in isostructural compounds is effective for tuning the AFM ground state and enhancing photomagnetic response. In the TM-thiophosphates, we show that even introducing 10\% of Ni ions into MnPS$_3$ -- a material characterized by minimal orbital angular momentum and inherently weak anisotropy -- dramatically enhances photomagnetism while leaving the ground state relatively untouched. Further tuning the chemical composition, we demonstrate precise control over the AFM ground state exchange and anisotropy, as well as the symmetry of the magnon mode. Control over these static and dynamic properties is essential from a device engineering perspective. While we sampled the full range of mixed compositions in this study, the TM-ion substitution fraction can be even more finely tuned to achieve desired parameters and performance. 

Understanding the microscopic mechanisms behind ultrafast photomagnetic processes remains a fundamental challenge~\cite{Penfold2023}. In light of this, mixing the compounds allows for a direct comparison of the photomagnetic response of Mn and Ni multiplet resonances within the same sample, which allows us to distill important considerations for the mechanisms of magnon generation. Our \textit{ab-initio} calculations demonstrate how spin conservation, orbital angular momentum and the degree of localization of multiplet resonances affect the magnitude of excited magnons. The $^3A_{1g}$ resonance in Ni$^{2+}$, in particular, drives the largest magnon response across the entire spectrum of mixed compounds due to its localized nature and the large unquenched orbital angular momentum of the excited state. The $^3T_{1g}$ multiplet, on the other hand, is delocalized across the ligands and may not be properly considered a \textit{d-d} excitation. Our observations also highlight the advantages of working with a low-symmetry system, in which multiplet resonances are non-degenerate. For comparison, in a high symmetry system such as NiO, the perfect cubic crystal field ensures that there are no \textit{d-d} excitonic transitions between 1 and 1.6~eV~\cite{propach1978farbe}.

These results naturally point to the broader exploration of TM dopants. In the TM-thiophosphate family, Co and Fe are compelling candidates, with the former hosting a partially filled $t_{2g}$ manifold and significant unquenched orbital angular momentum in the ground state compared to Ni and Mn, and the latter having the largest single-ion anisotropy of the family~\cite{leeIsing2016}. Altogether, the mixed TM-thiophosphates constitute a rich platform for the exploration and engineering of antiferromagnetic ground states and photomagnetic effects~\cite{li2022designing, yan2023first}. Beyond this family of materials, the principles we outline in this work can be applied to any collinear AFM where TM ions may be chemically substituted with ions of varying $d$-orbital occupation. As demonstrated here, intentional targeting of the multiplet resonances of dopants can significantly increase the efficacy and flexibility of photomagnetism in these materials.

\section*{Materials and methods}
\subsection{Sample growing and characterization}

Chemical vapor transport (CVT) technique is employed to grow bulk single crystals of (Mn$_x$Ni$_{1-x}$)PS$_3$ series of compounds following the synthesis method reported by Susner et al. with slight modification \cite{Susner2017}. Halogens (I$_2$ or Br$_2$) were employed as the transport agent in the synthetic process. Stoichiometric amounts of manganese (Mn) powder, nickel (Ni) powder, phosphorus (P) lump and sulfur (S) were sealed in an evacuated quartz tube (ampoule) with an inner pressure in the range of $10^{-5}$ to $10^{-6}$ Torr. In addition to the elemental constituents (Mn, Ni, P and S), a small amount (2 mg/cc) of iodine spheres were added inside the ampoule to act as the transport agent. The sealed ampoules were then subjected to a two-zone horizontal tube furnace.

Initially, the source zone was kept at 650$^{\circ}$C and the growth zone at 750$^{\circ}$C for 48 hours. This arrangement was made to allow the constituents to react completely and prevent the back transport or formation of undesired additional phases. After that, the temperature of the source zone gradually increased to 750$^{\circ}$C and that of the growth zone was lowered to 700$^{\circ}$C. This arrangement lasted for another 120 hours. Finally, the temperatures of both the zones were lowered down, and the plate-like crystals were subsequently obtained from the ampoule for further characterizations and experiments.

The characterization of the samples (composition, homogeneity) was performed by energy-dispersive X-ray spectroscopy (EDS). The EDS measurements were carried out on an Ultra-high Resolution Scanning Electron Microscope (SEM) SU8020 (Hitachi) equipped with a X-MaxN (Oxford) Silicon Drift Detector (SDD). Selected single crystals were glued to a carbon tape and analyzed by semiquantitative EDS at a 20 kV acceleration voltage. The details are presented in Supplementary Sec. 1.

\subsection{Magnetometry characterization}
The magnetometry measurements were performed using the Vibrating Sample Magnetometer (VSM) of a Physical Property Measurement System (PPMS) of Quantum Design. The pick-up coil was placed in the (vertical) constant magnetic field produced by the superconducting solenoid; one measures only the component of the magnetization $M$ along the field $H$ direction. For measuring parallel and perpendicular field configurations, the sample was taken out of the holder and remounted.

\subsection{Magneto-optical measurements}
Data shown in Fig.\,1b, Fig.\,4 of the main text are performed with a Spectra Physics Spitfire Ace system, operating at a repetition rate 1~kHz. The Spitfire Ace amplifies femtosecond pulses from a mode-locked MaiTai Ti:Sapphire seed laser (photon energy of 1.55~eV, pulse duration of 130~fs). A part of the generated beam is sent to a TOPAS-Prime collinear OPA, which generates pulses with tunable wavelengths from 0.06 - 6.56~eV, and is used as a pump. The remaining power, separated by the beamsplitter, is used as a probe. The pump is chopped at 500~Hz and focused to a spot size of $\approx 200~\upmu$m; the probe pulse is focused to a spot size of $\approx 120~\upmu$m.

The mixed compounds measured here were exfoliated to produce a smooth surface. The samples are bulk-like, with an estimated thickness between 100~$\upmu$m to 1~mm.
\\
\\
Data shown in Fig.\,1c of the main text (optical absorbance) are performed by using a halogen light source (Ocean Optics HL-2000) with a spectrum from 400~nm to 1600~nm. Thin NiPS$_3$ (MnPS$_3$) flakes of thickness $\approx$10~$\upmu$m and size $\approx 200~\upmu$m are prepared and put onto a CaF$_2$ substrate. The light is focused to a spot size of $<10~\upmu$m on the sample by an objective, and the transmitted light is collected by another objective, and focused onto a grating spectrometer, where an infrared (visible) camera is used to detect the transmitted intensity. 
\\
\\
Data shown in Fig.\,2 of the main text are performed with the PHAROS BURST laser system, which is an oscillator operating at 76~MHz, with a Yb:KGW crystal acting as an active medium. A regenerative amplifier (RA) and pulse-picker then allows the user to change the repetition rate without affecting the internal operation. We operate the PHAROS BURST RA at 75~KHz with 267~$\upmu$J of pulse energy, which is further pulse-picked down to 5~kHz for experiments to avoid excessive average-power heating of the sample. The probe pulses have a pulse duration of 170~fs with a photon energy of 1.2~eV. A percentage of the power of PHAROS pulses are split and used to see the ORPHEUS-HP optical parametric amplifier (OPA), which generates tunable pump wavelengths from 0.8 - 3~eV with a bandwidth of 20-30~meV, depending on the pump photon energy. The pump is chopped at 2.5~kHz, and focused to a spot size of $(94\pm15 \times 78\pm11)~\upmu$m full-width-at-half-maximum (FWHM) on the sample surface. The remaining power is used for the probe. To maximize the second harmonic generation signal, the probe pulses are focused tightly to a spot size of $\approx$50~$\upmu$m.

The mixed compounds measured here were exfoliated to produce a smooth surface. The samples are bulk-like, with an estimated thickness $\sim 100~\upmu$m.

\begin{acknowledgments}
The authors thank B.\,A.\,Ivanov for fruitful discussions, and K.\,Saeedi Ilkhchy and C.\,Berkhout for their technical support. We are grateful to Prof.~Hari Shrikant for insightful discussions. This work is supported by ERC grant 101078206 ASTRAL; programme “Materials for the Quantum Age” (QuMat, registration number 024.005.006), which is part of the Gravitation programme financed by the Dutch Ministry of Education, Culture and Science (OCW); the European Union Horizon 2020 and innovation program under the European Research Council ERC grant agreement no. 856538 (3D-MAGiC), and the European Research Council ERC grant agreement no. 101054664 273 (SPARTACUS). M.\,Na acknowledges the support of the Natural Sciences and Engineering Research Council of Canada (NSERC) PDF fellowship.

This work was authored in part by the National Laboratory of the Rockies for the U.S. Department of Energy (DOE) under Contract No. DE-AC36-08GO28308. Funding was provided by the Computational Chemical Sciences program within the Office of Basic Energy Sciences, U.S. Department of Energy. S.A. acknowledges the use of the National Energy Research Scientific Computing Center, under Contract No. DE-AC02-05CH11231 using NERSC award BES-ERCAP0021783 and S.A. also acknowledges that a portion of the research was performed using computational resources sponsored by the Department of Energy's Office of Energy Efficiency and Renewable Energy and located at the National Laboratory of the Rockies. The views expressed in the article do not necessarily represent the views of the DOE or the U.S. Government. The U.S. Government retains and the publisher, by accepting the article for publication, acknowledges that the U.S. Government retains a nonexclusive, paid-up, irrevocable, worldwide license to publish or reproduce the published form of this work, or allow others to do so, for U.S. Government purposes. 
\end{acknowledgments}

\section*{Contributions}
D.A., A.C., and M.N conceived the project. D.A., Th.R. and A.V.K. supervised the study. V.R., M.N. and D.A. performed the static and pump-probe experiments. D.P., M.V.S. and S.A. provided the theoretical model. A.C. provided the samples. E.K. and A.I provided the EDS characterization, A.V. provided the magnetometry. K.M., V.R., M.N., and P.C. performed optical absorption experiments. D.K., P.K., contributed to discussion and treatment of the experimental results. M.N., V.R., S.A., and D.A. wrote original draft with feedback from all the co-authors.

\section*{Competing interests}

The authors declare no competing interests.

\section*{Data availability}

All data supporting the findings of this paper are available from the corresponding authors upon request.

\newpage
\section*{References}
\bibliographystyle{unsrtnat} 
\bibliography{citations}
\clearpage
\newpage
\section*{Extended Data}
\renewcommand{\figurename}{\textbf{Extended Data Fig.}}
\setcounter{figure}{0}

\begin{figure*}[h!]
    \centering
    \includegraphics[width = 1\textwidth]{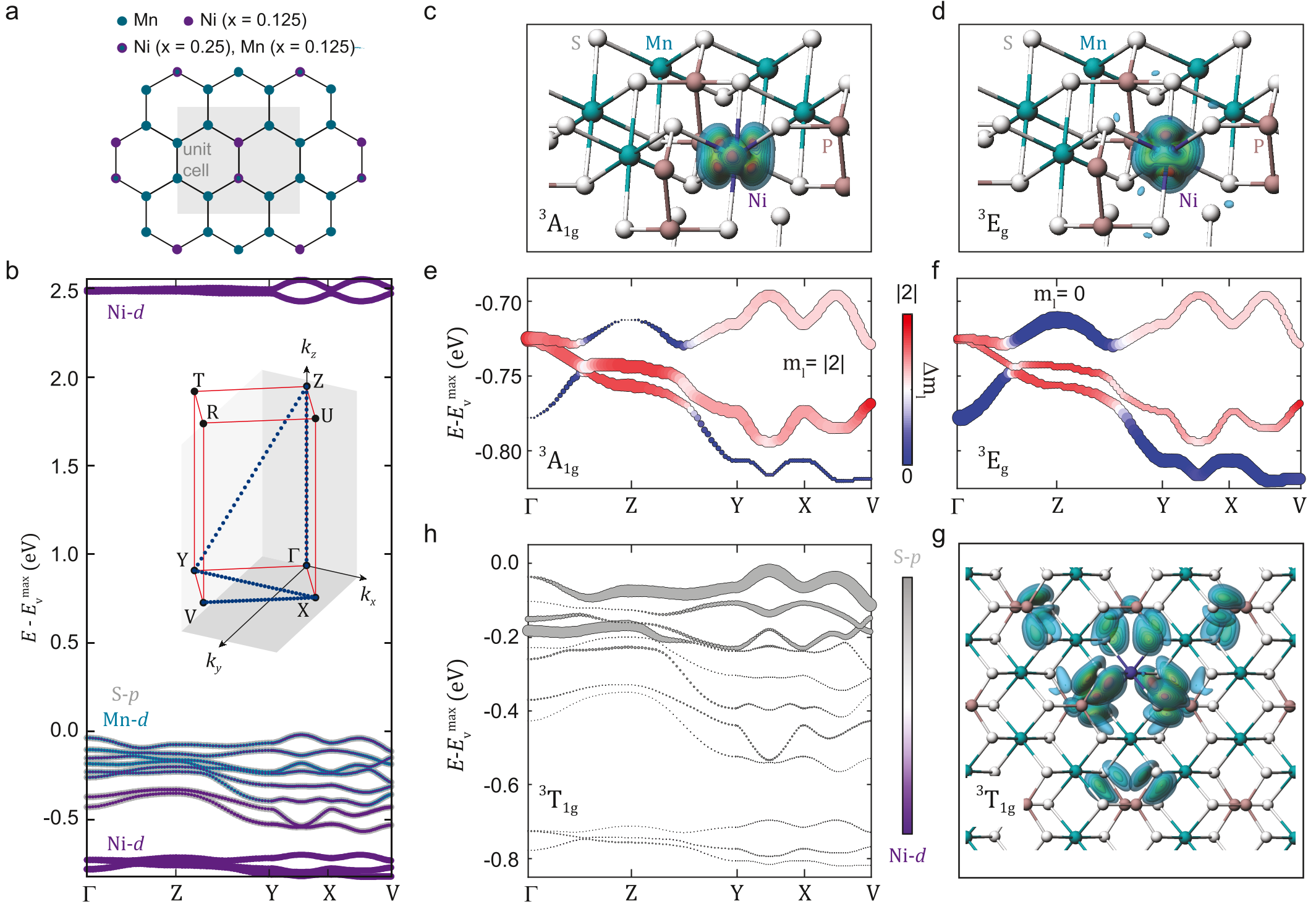} 
    \caption{\textbf{Ab-initio calculation of Ni$^{2+}$ multiplets.} Excitonic wavefunctions are computed in the framework of QS$G\hat{W}$~\cite{Cunningham2023} theory for the mixed compounds Mn$_{1-x}$Ni$_{x}$PS$_3$, with $x = 0.25$ (details in Supplementary Sec. 6). \textbf{(a)} The in-plane distribution of Mn and Ni ions for the $x = 0.125$ and $x = 0.25$ alloys. \textbf{(b)} The electronic band structure of Mn$_{1-x}$Ni$_{x}$PS$_3$, for $x = 0.25$. The colors of the markers indicate the weight of S-$p$, Mn-$d$, Ni-$d$ orbitals, respectively. The Brillouin zone and high-symmetry points are shown in the inset, with the $k$-path denoted in blue. \textbf{(c,d,g)} The wavefunction of the excitonic-hole for the $^3A_{1g}$, $^3E_{g}$ and  $^3T_{1g}$ excitations. The Ni and Mn ions are shown in purple and turquoise, respectively. S and P are shown in gray and brown. The $^3A_{1g}$ and $^3E_g$ excitations are localized to the Ni-ion, while the $^3T_{1g}$ state is delocalized over multiple hexagonal plaquette. \textbf{(e,f,h)} Electronic band structure plotted along $k$-path in the inset of b for the $^3A_{1g}$, $^3E_{g}$ and  $^3T_{1g}$ excitations. Marker size indicates the weight of the excitonic hole $c_{n, \mathbf{k}}$, raised to the power of 4 for visualization purposes. Projection show that the $^3A_{1g}$ ($^3E_{g}$) excitation is projected onto the $m_l = |2|\,(0)$ Ni-$d$ states, while the $^3T_{1g}$ excitation is delocalized over the S-$p$ valence bands.}
    \label{Fig: ext}
\end{figure*}

\end{document}